\documentclass[5p,twocolumn]{elsarticle}

\usepackage{graphicx}
\usepackage{dcolumn}
\usepackage{pifont}
\usepackage{bm}
\usepackage{multirow}
\usepackage{amsmath}
\usepackage{float}
\usepackage{txfonts}
\usepackage[version=3]{mhchem}

\usepackage[usenames,dvipsnames]{xcolor}
\definecolor{blue}{RGB}{0,0,225}
\definecolor{cream}{RGB}{222,217,201}
\definecolor{red}{RGB}{225,0,0}

\journal{arXiv}

\begin{document}
\title{Ab Initio Thermodynamic Study of \ce{PbI2} and \ce{CH3NH3PbI3} Surfaces in Reaction with \ce{CH3NH2} Gas for Perovskite Solar Cells}

\author[kimuniv-m]{Yun-Sim Kim}
\author[kimuniv-m]{Chol-Hyok Ri}
\author[kimuniv-m]{Yun-Hyok Kye}
\author[kimuniv-m]{Un-Gi Jong}
\author[kimuniv-m]{Chol-Jun Yu\corref{cor}}
\cortext[cor]{Corresponding author}
\ead{cj.yu@ryongnamsan.edu.kp}

\address[kimuniv-m]{Chair of Computational Materials Design (CMD), Faculty of Materials Science, Kim Il Sung University, Ryongnam-Dong, Taesong District, Pyongyang, PO Box 76, Democratic People's Republic of Korea}

\begin{abstract}
Hybrid organic-inorganic halide perovskites for photovoltaics have attracted research interest due to their unique material properties, but suffered from poor material stability.
In this work, we investigated the surface phase diagrams of \ce{PbI2} and cubic \ce{CH3NH3PbI3} (\ce{MAPbI3}) for a better understanding of precursor effect on perovskite synthesis via solid-gas reaction, by density functional theory calculations combined with thermodynamics.
Using the devised slab models of \ce{PbI2}(001) and \ce{MAPbI3}(100), (110) and (111) surfaces with perfect and various vacant defect terminations, we calculated their formation energies and adsorption energies of \ce{CH3NH2} molecule on the \ce{PbI2}(001) surfaces under different synthesis conditions of temperature, pressure and pH via chemical potentials of species.
Our calculations revealed that the adsorption can be facilitated by including HI or \ce{NH4I} molecule by dissociation of this additive and formation of \ce{CH3NH3+} cation or \ce{NH3}-\ce{CH3NH3+} complex, which is beneficial for conversion of \ce{PbI2} to \ce{MAPbI3} via solid-gas reaction.
Furthermore, we found that among different perovskite \ce{MAPbI3} surfaces, the MA-terminated (110) and MAI-terminated (100) surfaces are placed on the thermodynamically stable region of chemical potentials at pH values of 1 and 6, being agreed well with the experimental findings.
We believe this work gives a fundamental understanding of solid-gas reaction for high-crystallinity perovskite synthesis towards perovskite solar cells with improved stability.
\end{abstract}

\begin{keyword}
Methylammonium lead triiodide \sep Lead iodide \sep Methylamine \sep Surface \sep Phase diagram \sep Ab initio thermodynamics
\end{keyword}
\maketitle

\section{Introduction}
In the past decade, perovskite solar cells (PCSs) utilizing halide perovskites as photoabsorbers have attracted significant attention in photovoltaic community~\cite{Kojima09jacs,Saliba118s,Park20aem,McDonald19nanomat}.
Compared with other solar cells, the power conversion efficiency of PSCs has been rapidly growing, so that the latest conversion efficiency has been already certified to be 25.5\%~\cite{NREL}, being over the record of single crystalline Si solar cells.
Moreover, the fabrication cost is quite low due to the abundance of raw materials and facile synthesis method, highlighting their industrial potentiality and practicality~\cite{Li18nrm,Wang20natcom,Zhang20natcom}.
However, the long-term stability of PSCs was found to be poor~\cite{Lanzetta21nc,Berhe16ees,Shira16jap,Bryant16ees,Emilio16ees}, representing one of the challenges hindering their commercial utilization~\cite{Wang16semsc,Urbina20jpe}.
This is originated from constituent materials such as perovskite photoabsorbers and electron/hole transporting materials (E/HTMs) as well as device architecture.
To address this challenging issue, there have been developed numerous technologies including materials composition engineering~\cite{Prakash18mte,Miyasaka20aem,Lau17ael} and interface engineering~\cite{Kim21aami,Duan20se,Yu20acs,Clark20ael}.

Advancing fabrication methods for halide perovskites is one of the key ways for enhancing stability of PSCs.
In this line, Long {\it et al.}~\cite{Long16natcom} developed a two-step nonstoichiometric acid-base reaction route to increase moisture resistivity, which includes a conversion from the precursor \ce{HPbI3} to the perovskite \ce{CH3NH3PbI3} (methylammonium lead tri-iodide; \ce{MAPbI3}) using excess methylamine (\ce{CH3NH2}; MA$^{0}$) gas.
Liu {\it et al.}~\cite{Liu18natcom} introduced a simple method to fabricate perovskite films with over 1$\mu$m thickness via a fast solid-gas reaction of chlorine-incorporated hydrogen lead triiodide \ce{HPbI3}(Cl) and MA$^{0}$ gas.
Pang {\it et al.}~\cite{Pang16jacs} also reported the MA$^{0}$ gas-based fabrication method to induce the transformative evolution of ultrasmooth and full-coverage perovskite thin films from rough and partial-coverage \ce{HPbI3} thin films.
Therefore, the MA$^{0}$ gas treatment is regarded as an effectual process to produce a high-quality perovskite layer and improve its material stability.

In the meantime, as another approach to solve the long-term stability issue of PSCs, HTM-free carbon electrode-based PSCs have been envisioned, based on the consideration that the organic HTM, e.g., Spiro-OMeTAD, is one of main causes for chemical degradation of PSCs~\cite{Ko19semsc,Wei18se,LLiu15jacs,YRong14jpcl,Stranks13sci}.
In these HTM-free PSCs, perovskite layer serves as hole transport layer as well as the photoabsorber, and thus the quality of perovskite layer is crucial to the performance.
In addition to high-quality perovskite layer, MA$^{0}$ gas treating also improves the surface smoothness of perovskite layer on the thick \ce{TiO2} mesoporous scaffold, which is directly associated with a better contact between the perovskite and carbon electrode layers in the carbon-based PSCs~\cite{Zong16aciee}.
Wei {\it et al.}~\cite{Wei18se} introduced the MA$^{0}$ gas treatment process as the second step for converting precursor materials to the perovskite \ce{MAPbI3} layer for printable carbon-based PSCs, demonstrating that among different precursors \ce{PbI2}+HI and \ce{HPbI3}+\ce{PbI2}+\ce{NH4I} yield a fast and perfect perovskite conversion.
In spite of the extensive experimental evidences for a crucial role of MA$^{0}$ gas in improving the quality of perovskite layer, no theoretical modeling has yet performed, remaining the underlying mechanism unknown.

In this work, we investigate the initial converting process of \ce{PbI2} precursor to perovskite \ce{MAPbI3} with MA$^{0}$ gas introduction from first-principles.
We make modeling for \ce{PbI2} surface and MA$^{0}$ molecule adsorption on that surface in the presence of \ce{HI} and \ce{NH4I}, and calculate adsorption energies under consideration of different synthesis conditions.
We also calculate the formation enthalpy of the low index surfaces of resultant \ce{MAPbI3}, including (100), (110) and (111) surfaces with different terminations, to investigate the effect of growth conditions, in particular, the partial pressure of MA$^{0}$ gas.
This work gives a comprehensive understanding of solid-gas reaction process for organic-inorganic hybrid perovskites and thus paves the way for fabricating high-performance PSCs.

\section{Methods}
All calculations within density functional theory (DFT) framework were performed by using the pseudopotential plane-wave method as implemented in the Quantum ESPRESSO (QE, version 6.2) package~\cite{QE}.
With additional consideration of scalar relativistic effect for heavy I and Pb atoms, the ultrasoft pseudopotentials of constituent atoms were constructed using the LD1 code and input files from the PSlibrary (version 1.0.0), where the electron configurations are Pb-$5d^{10}6s^26p^2$, I-$4d^{10}5s^25p^5$, C-$2s^22p^2$, H-$1s^1$, O-$2s^22p^4$ and N-$2s^22p^3$.
To account for the effect of van der Waals (vdW) interactions between the molecular moieties, we adopted the rev-vdW-DF2 exchange-correlation functional~\cite{rev14prb,vdw04prl,vdw07prb}.
We set kinetic cutoff energies as 70 and 700 Ry for the wave function and electron density, and special $k$-point meshes as $6\times6\times6$ and $2\times2\times1$ for bulk and surface calculations, respectively.
With these parameters, the total energy of bulk and formation energy of surface converged within 5 meV per atom and 5 mJ/m$^2$, respectively.
All atoms were relaxed until the atomic forces converged to $5\times10^{-4}$ Ry/Bohr, and for the lattice optimization, until the stress was less than 0.05 kbar.

As a preliminary step, we optimized the bulk unit cells of rhombohedral \ce{PbI2} with space group $P\bar{3}m1$ and pseudo-cubic \ce{MAPbI3} with space group $Pm\bar{3}m$.
Then, we built surface models by cleaving the optimized bulk unit cells along the selected crystallographic directions, considering different terminations.
For \ce{PbI2}, we selected (001) surface, which is regarded as the most stable surface with the lowest surface formation energy, with Pb and I terminations, and made supercells using inversion-symmetric slab models consisted of nine atomic layers and 15 \AA-thick vacuum layer.
When increasing the atomic layers from nine to eleven, the surface energy was found to increase by only 0.01 J/m$^2$.
To consider defective surfaces and different adsorbate coverage, the surface unit cells were set to $2\times2$ as well as $1\times1$ for these slab models.
For the defective surfaces, we created vacancy defects such as $V_{\ce{I}}$ and $V_{\ce{Pb}}$ in the vicinity of top surface with I and Pb terminations, respectively.
Three atomic layers from top layer in both sides were allowed to relax, while the middle three layers were fixed at their bulk positions.
Then the surface free energy is calculated as follows:
\begin{equation}
\gamma=\frac{1}{2A}\Big( E_{\text{slab}}\big[(\ce{PbI2})_\alpha\ce{Pb}_\beta\ce{I}_\delta\big]-\alpha\mu_{\ce{PbI2}} - \beta\mu_{\ce{Pb}} - \delta\mu_{\ce{I}} \Big) \label{eq1}
\end{equation}
\begin{equation}
\begin{gathered}
\hspace{-60pt}\mu_{\ce{PbI2}} = E_{\ce{PbI2}}^{\text{rhom}} + \Delta\mu_{\ce{PbI2}},\\
~\mu_{\ce{Pb}} = E_{\ce{Pb}}^{\text{fcc}} + \Delta\mu_{\ce{Pb}}, 
~\mu_{\ce{I}} = E_{\ce{I}}^{\text{orth}} + \Delta\mu_{\ce{I}}. \label{eq2}
\end{gathered}
\end{equation}
where $A$ is the surface area, $E^{\text{phase}}_{\text{comp}}$ is the total energy of ``compound'' in ``phase'', and $\mu_{\text{spec}}$ is the chemical potential of ``species''.
We consider thermodynamic constraints for the chemical potentials of Pb, I and \ce{PbI2} based on the fact that the bulk \ce{PbI2} material is in equilibrium state, as follows,
\begin{equation}
\Delta\mu_{\ce{PbI2}} = 0, ~~\Delta\mu_{\ce{Pb}} + 2\Delta\mu_{\ce{I}} = \Delta H_{\ce{PbI2}},
\label{eq3}
\end{equation}
where $\Delta H_{\ce{PbI2}} = E_{\ce{PbI2}}^{\text{rhom}} - E_{\ce{Pb}}^{\text{fcc}} - 2E_{\ce{I}}^{\text{orth}}$ is the formation enthalpy of \ce{PbI2} bulk in rhombohedral phase.
Here, Pb bulk is in face-centered cubic (fcc) phase with space group $Fm\bar{3}m$, and I bulk in orthorhombic (orth) phase with space group $Cmca$.

Given the relaxed clean surface models, we studied adsorption of MA$^0$ molecule on the surfaces in the presence of HI or \ce{NH4I} molecule.
After rough prediction of adsorption configuration, atomic relaxations were again performed and the adsorption energies were calculated using the total energies of supercells for the molecule-adsorbed surface, clean surface, and isolated molecule, respectively.
The molecule was simulated using the cubic supercell with a cell length of 15 \AA.

To investigate the effect of growth condition on the formation of \ce{MAPbI3}, we also considered its surface with only low-indexes such as (100), (110) and (111) with different terminations using the slab models with $1\times1$ surface unit cells, nine atomic layers and 15 \AA~thick vacuum layer.
The surface formation energy per area is calculated as follows,
\begin{multline}
\gamma=\frac{1}{2A}\Big( E_{\text{slab}}\big[(\ce{MAPbI3})_\alpha\ce{Pb}_\beta\ce{I}_\delta\ce{MA}^0_\eta\ce{H}_\zeta\big] - \\
\hspace{50pt} - \alpha\mu_{\ce{MAPbI3}} - \beta\mu_{\ce{Pb}} - \delta\mu_{\ce{I}} - \eta\mu_{\ce{MA^0}} - \zeta\mu_{\ce{H}}\Big), \label{eq4} 
\end{multline}
\begin{equation}
\begin{gathered}
\hspace{-25pt}\mu_{\ce{MAPbI3}} = E_{\ce{MAPbI3}}^{\text{cub}} + \Delta\mu_{\ce{MAPbI3}},\\
\hspace{28pt}~\mu_{\ce{MA^0}} = E_{\ce{MA^0}}^{\text{gas}} + \Delta\mu_{\ce{MA^0}}, 
~\mu_{\ce{H}} = \frac{1}{2}E_{\ce{H2}}^{\text{gas}} + \Delta\mu_{\ce{H}}.\label{eq5}
\end{gathered}
\end{equation}
Assuming that $\Delta\mu_{\ce{MAPbI3}}=0$ and the compounds are under equilibrium, the thermodynamically stable condition for \ce{MAPbI3} is expressed as,
\begin{equation}
\Delta\mu_{\ce{Pb}} + 3\Delta\mu_{\ce{I}} + \Delta\mu_{\ce{MA^0}} + \Delta\mu_{\ce{H}} = \Delta H_{\ce{MAPbI3}},
\label{eq6}
\end{equation}
where $\Delta H_{\ce{MAPbI3}} = E_{\ce{MAPbI3}}^{\text{cub}} - E_{\ce{Pb}}^{\text{fcc}} - 3E_{\ce{I}}^{\text{orth}} - E_{\ce{MA^0}}^{\text{gas}} - \frac{1}{2}E_{\ce{H2}}^{\text{gas}}$ is the formation enthalpy of \ce{MAPbI3} bulk in cubic phase.
Here, $\Delta\mu<0$ must be satisfied to prevent the formation of simple substance in the conversion.
Also, the competing binary or ternary compounds should not be formed, giving the constraints,
\begin{eqnarray}
\hspace{30pt}\Delta\mu_{\ce{Pb}} + 2\Delta\mu_{\ce{I}} \leq \Delta H_{\ce{PbI2}}, \hspace{1.05cm}
\label{eq7}\\
\hspace{30pt}\Delta\mu_{\ce{MA^0}} + \Delta\mu_{\ce{I}} +\Delta\mu_{\ce{H}} \leq \Delta H_{\ce{MAI}},
\label{eq8}
\end{eqnarray}
where $\Delta H_{\ce{MAI}} = E_{\ce{MAI}}^{\text{cub}}-E_{\ce{MA^0}}^{\text{gas}}-E_{\ce{I}}^{\text{orth}}$ is the formation enthalpy of MAI bulk in cubic phase with space group $Pm$.
To prevent the dissociation of \ce{MAPbI3} into its constituents of \ce{PbI2} and MAI, the following constraint must be satisfied,
\begin{equation}
\Delta H_{\ce{MAPbI3}} - \Delta H_{\ce{MAI}} \leq \Delta\mu_{\ce{Pb}} + 2\Delta\mu_{\ce{I}} \leq \Delta H_{\ce{PbI2}}.
\label{eq9}
\end{equation}
The calculated total energies and formation enthalpies of all the compounds under study, with their phases, are given in Table S1 in the Supporting Information.

The chemical potential of MA$^0$ varies according to the partial pressure $P$ and temperature $T$ as follows,
\begin{equation}
\Delta\mu_{\ce{MA^0}}(T,~P_{\ce{MA^0}})=\Delta\mu_{\ce{MA^0}}(T,~P^\circ)+k_{\text{B}}T\ln\frac{P_{\ce{MA^0}}}{P^\circ},
\label{eq10}
\end{equation}
where $P^\circ$ is the atmospheric pressure and $k_{\text{B}}$ is the Boltzmann constant.
We extracted the value of $\Delta\mu_{\ce{MA^0}}(T,~P^\circ)$ from the experimental data~\cite{Lide07crc} as follows,
\begin{equation}
\Delta\mu_{\ce{MA^0}}(T,~P^\circ)\approx[H^\circ(T)-H^\circ(T_{\text{r}})]+[H^\circ(T_{\text{r}})-H^\circ(0)]-TS^\circ(T),
\label{eq11}
\end{equation}
where $T_{\text{r}}=298.15$ K is the reference temperature, and $H^\circ$ and $S^\circ$ are the enthalpy and entropy measured at $P^\circ$~\cite{Hong16pccp}.

The presence of HI solution allows us to consider pH condition as the chemical potential of hydrogen $\Delta\mu_{\ce{H}}$, as it is an inverse linear function of the pH value~\cite{Watcharatharapong18ne}.
When $\Delta\mu_{\ce{H}}$ reaches its maximum ($\Delta\mu_{\ce{H}}^{\text{max}}=\frac{1}{2}E_{\ce{H2}}^{\text{gas}}$) or minimum ($\Delta\mu_{\ce{H}}^{\text{min}}=E_{\ce{H2O}}^{\text{gas}}-E_{\ce{OH}}^{\text{gas}}$) according to $\mu_{\ce{H2O}}=\mu_{\ce{H+}}+\mu_{\ce{OH-}}$, where $\mu_{\ce{H+}}=\mu_{\ce{H}}$ for simplification, the pH value will be 0 (acidic) or 14 (basic).
Then, the pH value can be estimated by linear interpolation as $\text{pH}=14(\Delta\mu_{\ce{H}}-\Delta\mu_{\ce{H}}^{\text{max}})/(\Delta\mu_{\ce{H}}^{\text{min}}-\Delta\mu_{\ce{H}}^{\text{max}})$~\cite{Watcharatharapong18ne}.

\begin{figure}[!b]
\centering
\includegraphics[clip=true,scale=0.09]{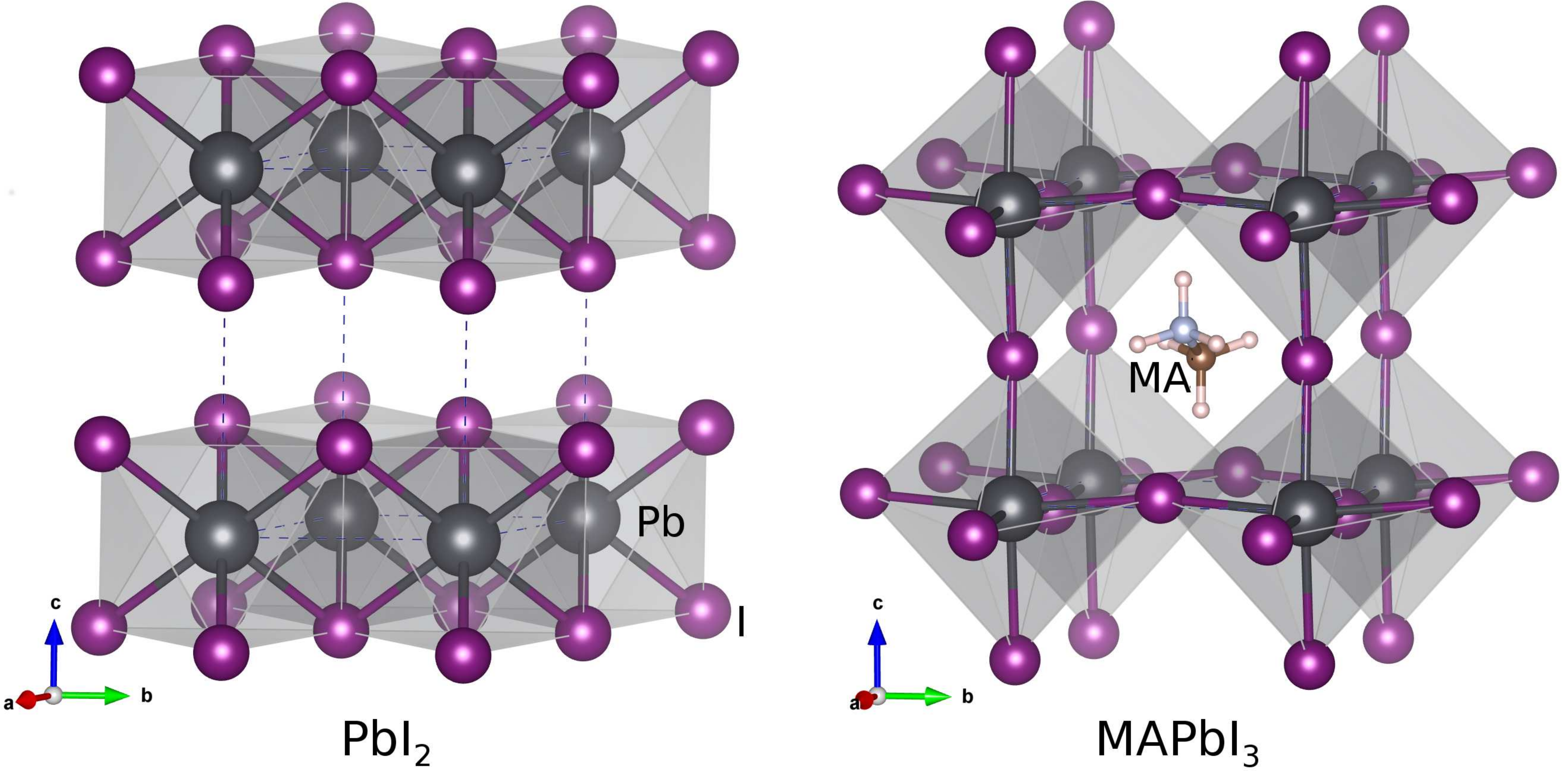}
\caption{Polyhedral view of optimized bulk unit cells of rhombohedral \ce{PbI2} (left) and cubic \ce{CH3NH3PbI3} (right).
Black, purple, brown, blue and white spheres denote Pb, I, C, N and H atoms, respectively.}
\label{fig1}
\end{figure}

\section{Results and discussion}
\subsection{Surface phase diagram of \ce{PbI2} surface}
We first calculated the lattice constants of bulk unit cells of \ce{PbI2} and \ce{MAPbI3} by optimization.
The obtained lattice constants were $a=b=4.572$ \AA, $c=7.018$ \AA~for \ce{PbI2} and 6.337 \AA~for \ce{MAPbI3}, which agree well with their experimental values of $a=b=4.558$ \AA, $c=6.986$ \AA~\cite{Palosz90jpcm} and $a=6.329$ \AA~\cite{Poglitsch87jpc}, respectively (Figure~\ref{fig1}).
We also performed atomic relaxations of the isolated \ce{CH3NH2} molecule, for which the bond lengths are 1.101, 1.020 and 1.475 \AA~for C$-$H, N$-$H and C$-$N bonds in good agreement with the previous theoretical and experimental values of 1.093, 1.011 and 1.474 \AA~\cite{Cheng09jms} (Figure S1).
These indicate that the calculations undertaken here could provide reasonable description for the systems under study.

\begin{figure}[!b]
\centering
\includegraphics[clip=true,scale=0.51]{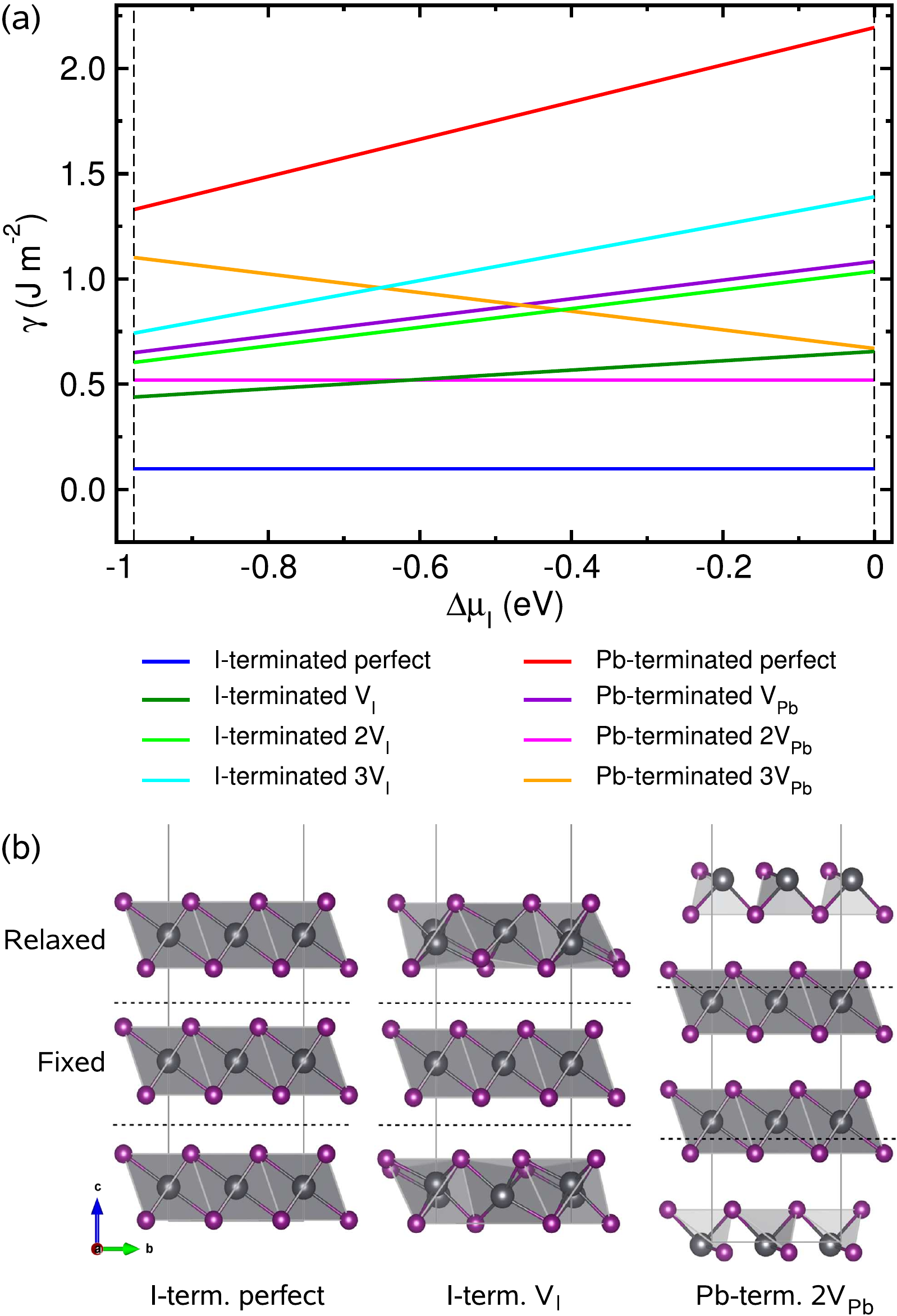}
\caption{(a) Surface formation energies of eight \ce{PbI2}(001) surfaces with I- and Pb-terminations and vacancy defects as functions of iodine chemical potential.
Two dashed vertical lines denote the lower ($\Delta\mu_\text{I}=-0.98$ eV) and upper (0.0 eV) limits of the iodine chemical potential.
(b) Polyhedral view of the lowest formation energy surfaces with $2\times2$ supercell; I-terminated perfect (left), I-terminated $V_{\ce{I}}$ (middle) and Pb-terminated $2V_{\ce{Pb}}$ (right) defective surfaces.}
\label{fig2}
\end{figure}

We then performed surface relaxations of \ce{PbI2}(001) surface slabs with 2$\times$2 surface supercell and different terminations.
We considered the two different terminations, i.e., I- and Pb-terminations, for which perfect and defective surfaces were constructed.
The stoichiometric (perfect) I-terminated \ce{PbI2}(001) surface is characterized by type-2 according to the Tasker's classification scheme~\cite{Tasker79}, indicating that there is no net dipole moment perpendicular to the surface and therefore it can be easily formed from bulk as a stable surface without any reconstruction under vacuum condition.
On the contrary, Pb-termination invokes the surface dipole moment, thereby making it difficult to create the surface.
When the surface formation occurs under ambient condition in contact with its reservoirs such as iodine and lead crystal, various surface reconstructions should be considered due to the interaction with them.
Therefore, it is natural that the defective \ce{PbI2}(001) surfaces are expected to be formed at a certain condition given by the chemical potential of iodine.
We considered one, two, and three I vacancies for I-terminated perfect surface, and the same number of Pb vacancies for Pb-terminated surface.
In total, eight different models for the \ce{PbI2}(001) surface in contact with iodine crystal using $2\times2$ surface supercells were constructed in this work (see Figures S2 and S3 in the Supporting Information).

We computed the formation energies of these eight surfaces by applying Eqs.~\ref{eq1} and~\ref{eq2}.
Figure~\ref{fig2}(a) presents the calculated phase diagram of the \ce{PbI2}(001) surface as a function of the iodine chemical potential.
As expected above, the I-terminated perfect surface was found to have the lowest formation energy of 0.09 J/m$^2$ among the eight surfaces.
At lower limit of iodine chemical potential under the I-poor/Pb-rich condition, which is $-0.98$ eV estimated from the bulk \ce{PbI2} formation energy of $-1.95$ eV, the I-terminated $V_{\ce{I}}$ defective surface has the second lowest formation energy of $\sim$0.45 J/m$^2$.
In fact, creating I vacancies on the perfect I-terminated surface induces surface dipole moment and remarkable surface relaxation, which get more pronounced as increasing the number of I vacancies (see Figure S2), resulting in the increase of formation energy.
On the other hand, under the I-rich/Pb-poor condition, the Pb-terminated 2$V_{\ce{Pb}}$ defective surface was shown to be favorable for the formation, which is related with the fact that removing two Pb atoms recovers \ce{PbI2} stoichiometry on the surface, thereby net-zero dipole moment.
Figure~\ref{fig2}(b) presents polyhedral view of these three relaxed surfaces, which will be used in further calculations for molecule adsorption.

\subsection{\ce{CH3NH2} adsorption on \ce{PbI2}(001) surface}
To simulate the initial process of \ce{MAPbI3} perovskite formation via solid-gas reaction between \ce{PbI2} solid and \ce{CH3NH2} gas~\cite{Liu18natcom,Pang16jacs,Wei18se}, we investigated adsorption of \ce{CH3NH2} (MA$^0$) molecule on the \ce{PbI2}(001) surface.
As aforementioned, the three types of surfaces with the lowest formation energies, i.e., I-terminated perfect, I-terminated $V_{\ce{I}}$ defect, and Pb-terminated 2$V_{\ce{Pb}}$ defect surfaces were used as substrates.
We also appended a HI or \ce{NH4I} molecule as an additional adsorbate, since HI or \ce{NH4I} gas could be added to facilitate the reaction in experiment~\cite{Wei18se}.
Two different surface cells of $1\times1$ and $2\times2$ were used for I-terminated perfect surface to consider different concentrations of adsorbate coverage.

For calculation of adsorption energy, we considered the enthalpy and entropy terms for MA$^0$ gas at finite temperature, while for solids only the DFT total energies were used because the both terms are in general $3\sim4$ order smaller than the internal energy difference.
For the MA$^0$ molecule adsorption, the adsorption energy can be calculated as follows,
\begin{equation}
E_{\text{ads}}=E_{\text{surf-MA}^0}-(E_{\text{surf}}+G_{\text{MA}^0})
\label{eq12}
\end{equation}
where $E_{\text{surf-MA}^0}$ and $E_{\text{surf}}$ are the total energies of supercells for the MA$^0$-adsorbed surface and clean surface, respectively.
The Gibbs free energy of MA$^0$ molecule in gas phase, $G_{\text{MA}^0}$, was obtained by applying Eqs.~\ref{eq10} and~\ref{eq11} using the DFT total energy as the internal energy and the experimental data for enthalpy and entropy terms from experimental data~\cite{Lide07crc}.
In the presence of \ce{HI} or \ce{NH4I}, the adsorption energy was evaluated as,
\begin{equation}
E_{\text{ads}}=E_{\ce{surf-HI}/\ce{NH4I-MA}^0}-(E_{\text{surf}}+E_{\ce{HI}/\ce{NH4I}}+G_{\text{MA}^0})
\label{eq13}
\end{equation}
where $E_{\ce{HI}/\ce{NH4I}}$ is the total energy of the isolated HI or \ce{NH4I} molecule in supercell, simulating the aqueous solution.
Table~\ref{tab1} presents the calculated adsorption energies of MA$^0$ molecule on the different types of \ce{PbI2}(001) surfaces without or with the presence of HI or \ce{NH4I} molecule at different temperatures of 0, 300 and 400 K.
\begin{table}[!th]
\small
\caption{\label{tab1}The calculated adsorption energies of \ce{CH3NH2} molecule on different types of \ce{PbI2}(001) surfaces without or with the presence of HI or \ce{NH4I} at different temperatures.}
\begin{tabular}{llccc}
\hline
& & \multicolumn{3}{c}{$E_{\text{ads}}$ (eV)} \\ 
\cline{3-5}
Surface type & Additive & 0 K & ~~300 K & ~~400 K  \\
\hline
1$\times$1 I-term. perf. & -- & $-$0.37  & ~~0.29 & ~~0.55 \\
 & HI & $-$1.08 & $-$0.42 & $-$0.16 \\
\hline
2$\times$2 I-term. perf. & -- & $-$0.03 & ~~0.64 & ~~0.90 \\
 & HI & $-$1.00  & $-$0.34 & $-$0.08 \\
& \ce{NH4I} & $-$0.84  & $-$0.18 & ~~0.08 \\
\hline
2$\times$2 I-term. $V_\ce{I}$ & -- & $-$0.63  & ~~0.03 & ~~0.29 \\
& HI & $-$1.67 & $-$1.01 & $-$0.75 \\
& \ce{NH4I} & $-$1.45  & $-$0.78 & $-$0.52 \\
\hline
2$\times$2 Pb-term. $2V_\ce{Pb}$ & -- & $-$0.65 & ~~0.02 & ~~0.25 \\
 & HI & $-$3.25  & $-$2.59 & $-$2.33 \\
& \ce{NH4I} & $-$3.36  & $-$2.70 & $-$2.43 \\
\hline
\end{tabular}	
\end{table}

The adsorption energy on the I-terminated perfect surface with $1\times1$ cell, i.e., with a coverage of 1ML, was calculated to be $-0.37$ eV at 0 K, indicating that the adsorption is exothermic.
As temperature increases, the adsorption energy changed from being negative to positive (0.29, 0.55 eV at 300, 400 K), implying that the adsorption became endothermic over room temperature.
The adsorption looks like being molecular through the weak hydrogen bond between H atom of amino group of MA$^0$ molecule and I atom of substrate surface with a H$-$I length of 3.17 \AA, for which some amount of electrons are shown to be transferred from the molecule to the surface (Figure S4).
When adding a HI molecule, the adsorption could occur more easily by seeing that the adsorption energy became lower as $-1.08$ eV at 0 K and negative even at higher temperatures ($-0.42$, $-0.16$ eV at 300, 400 K).
For comparison, we also calculated the adsorption energy of only HI molecule on the same surface, which was 0.16 eV.
During the adsorption, two-step chemical reactions were found to occur, i.e., dissociation of HI molecule and binding \ce{H+} ion to \ce{CH3NH2} molecule, forming \ce{CH3NH3+} cation (Figure S4), which can be a cause of lowering adsorption energy by adding HI molecule.
This is consistent with the experimental finding that the conversion is significantly enhanced in the presence of \ce{H+}/\ce{I-} or H-I group~\cite{Wei18se}.

\begin{figure*}[!th]
\centering
\includegraphics[clip=true,scale=0.17]{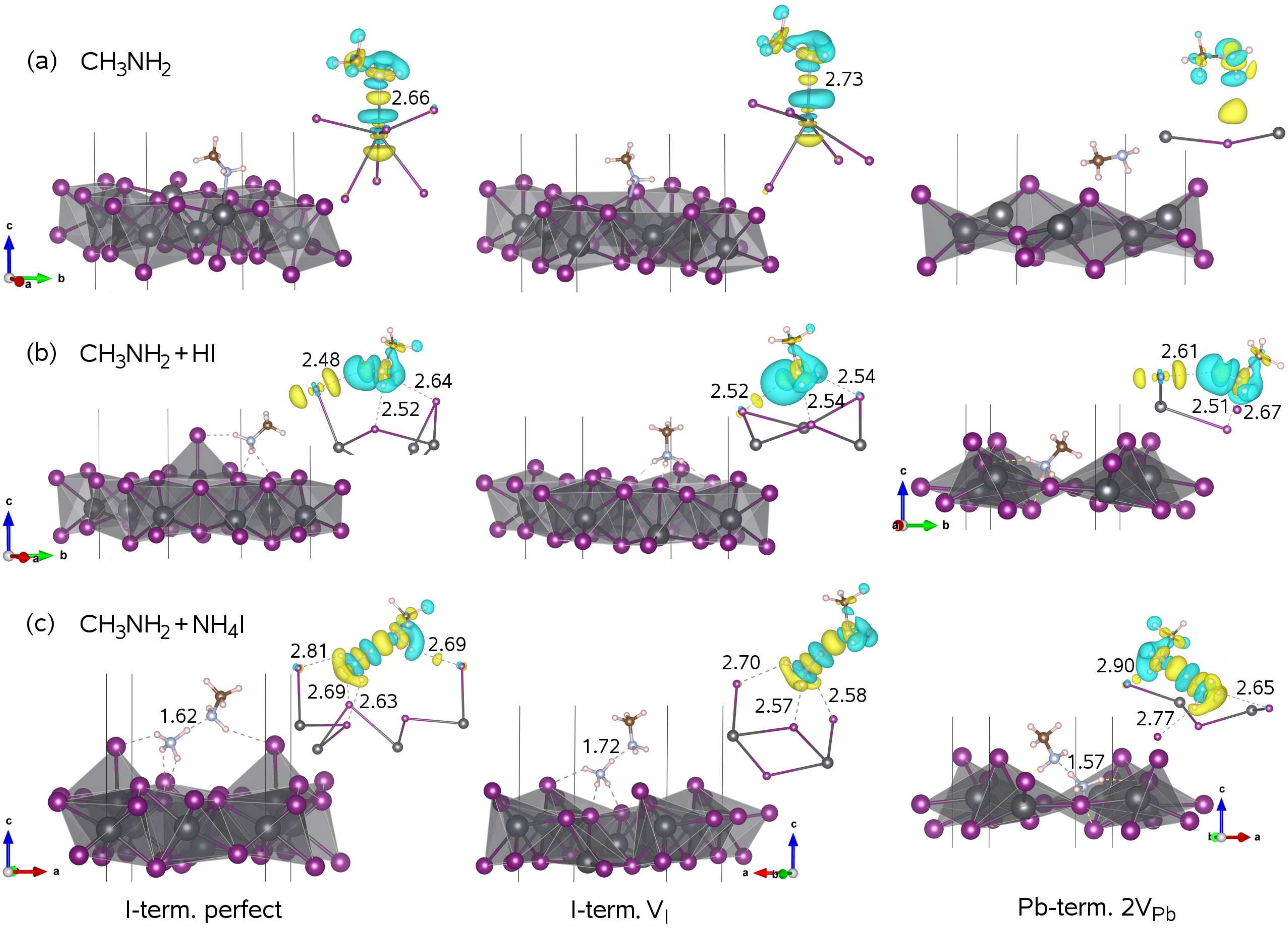}
\caption{Relaxed geometries of \ce{CH3NH2} molecule (a) without and with (b) HI or (c) \ce{NH4I} molecule adsorbed on \ce{PbI2}(001) I-terminated perfect (left panel), I-terminated $V_{\ce{I}}$ (middle) and Pb-terminated $V_{\ce{Pb}}$ defective surfaces with $2\times2$ surface cell.
Hydrogen bonds between H atoms of the adsorbate and I atoms of the surface are denoted by dashed lines with bond lengths, and N$-$Pb (top panel) and N$-$H (bottom panel) bond lengths are also denoted (unit: \AA).
The isosurface view of electronic charge difference occurred around adsorbate is given at the top-right corner in each case, where yellow (cyan) color denotes the electron accumulation (depletion) at the same value of 0.002 $|e|$/\AA$^3$.}
\label{fig3}
\end{figure*}
Then, we considered the adsorption of MA$^0$ molecule on the three types of \ce{PbI2}(001) surfaces with 2$\times$2 surface cells (0.25 ML coverage), selected from the surface formation energy evaluation.
The effect of HI and \ce{NH4I} inclusion was also considered.
Figure~\ref{fig3} shows the relaxed geometries of MA$^0$ molecule without and with the presence of HI or \ce{NH4I} molecule adsorbed on the \ce{PbI2}(001) I-terminated perfect, I-terminated $V_{\ce{I}}$ and Pb-terminated $2V_{\ce{Pb}}$ defective surfaces.
For the cases of only MA$^0$ molecule adsorptions, the adsorption energies were determined to be $-0.03$, $-0.63$ and $-0.65$ eV for the three types of surfaces in the same order to above text, indicating that the defective surfaces are more favorable to MA$^0$ adsorption than the perfect surface.
As shown in Figure~\ref{fig3}(a), the adsorptions occur through chemical bonding between N atom and surface Pb atom with N$-$Pb bond lengths of 2.66 and 2.73 \AA~for I-terminate perfect and defect surfaces, which is confirmed by electron transfer from adsorbate to the Pb atom upon adsorption.
For the Pb-terminated defective surface, no explicit binding was observed.
As increasing temperature, the adsorption energies became positive like 0.64, 0.03, 0.02 eV at 300 K, and 0.90, 0.29, 0.25 eV at 400 K for the three types of surfaces, respectively.

When adding a HI or \ce{NH4I} molecule, the adsorption energies were found to be lower for all the cases, indicating easier conversion by this additive.
For the I-terminated surfaces, the HI addition has slightly enhanced effect than \ce{NH4I} due to lower adsorption energies in the former cases ($-1.00$, $-1.67$ eV) than those in the latter cases ($-0.84$, $-1.45$ eV).
Meanwhile, \ce{NH4I} addition ($-3.36$ eV) is more favorable to MA$^0$ adsorption than HI one ($-3.25$ eV) for the Pb-terminated defective surface.
Moreover, it was found that the defective surfaces are preferable to the perfect surface, and the Pb-terminated $2V_{\ce{Pb}}$ defective surface is the most desirable for adsorption. 
To understand the underlying mechanism, we provide the analysis of optimized geometries with electronic charge redistribution.
As shown in Figure~\ref{fig3}(b) for the cases of MA$^0$+HI adsorption, HI was dissociated into \ce{H+} and \ce{I-} ions, of which the \ce{H+} cation was bonded with N atom to form \ce{CH3NH3+} cation and \ce{I-} anion was combined with the \ce{PbI,{$n$}} polyhedron.
For the I-terminated $V_{\ce{I}}$ surface, the dissociated \ce{I-} anion occupied the I-vacant site of the \ce{PbI5} polyhedron to recover the \ce{PbI6} octahedron, while for the perfect surface the abnormal \ce{PbI7} polyhedron was formed (Figure S5), thereby implying that the defective surface is preferable.
For the Pb-terminated $2V_{\ce{Pb}}$ surface, the additional \ce{I-} anion leads to forming a pair of face-sharing \ce{PbI5} square pyramids (Figure S5), which are connected each other via corner-sharing and thus are relatively stable.
The newly formed \ce{MA+} cation was strongly bound to the surface through hydrogen bonds with H$-$I bond lengths ranging from 2.48 to 2.67 \AA.

Similar findings were obtained for the \ce{NH4I} addition.
That is, the \ce{NH4I} was dissociated into \ce{NH4+} and \ce{I-} ions, which were bound with MA$^0$ through the weak hydrogen bond (bond length spans from 1.57 to 1.72 \AA) and \ce{PbI,{$n$}} respectively.
Since the dissociation of \ce{NH4I} is likely to occur more hardly than that of HI and the binding of \ce{NH4}$-$\ce{CH3NH2} is relatively weak, the effect of \ce{NH4I} is slightly weaker than HI.
For the case of I-terminated $V_{\ce{I}}$ surface, the \ce{I-} ion is not recovering \ce{PbI6} octahedron due to the strong interaction from \ce{NH4+} (Figure~\ref{fig3}(c) and Figure S6), resulting in slightly higher adsorption energy (Table~\ref{tab1}).
Meanwhile, similar configuration of substrate surface was found for the Pb-terminated defective surface, while enhanced interaction between adsorbate and substrate was observed, resulting in lowering the absorption energy compared with that by HI addition.
These findings indicate that both HI and \ce{NH4I} play a positive role in synthesizing \ce{MAPbI3} from \ce{PbI2} precursor by combining with the surface Pb atom with \ce{I-} and helping MA$^0$ gas adsorption on that with either \ce{H+} or \ce{NH4+} cation.
It is now confirmed that the higher crystallinity and superior morphology of \ce{MAPbI3} layer synthesized from \ce{NH4I}+\ce{PbI2} precursor, reported in Ref.~\cite{Wei18se}, is attributed to the slow conversion process including the release of \ce{NH3} gas.

\begin{figure}[!b]
\centering
\includegraphics[clip=true,scale=0.5]{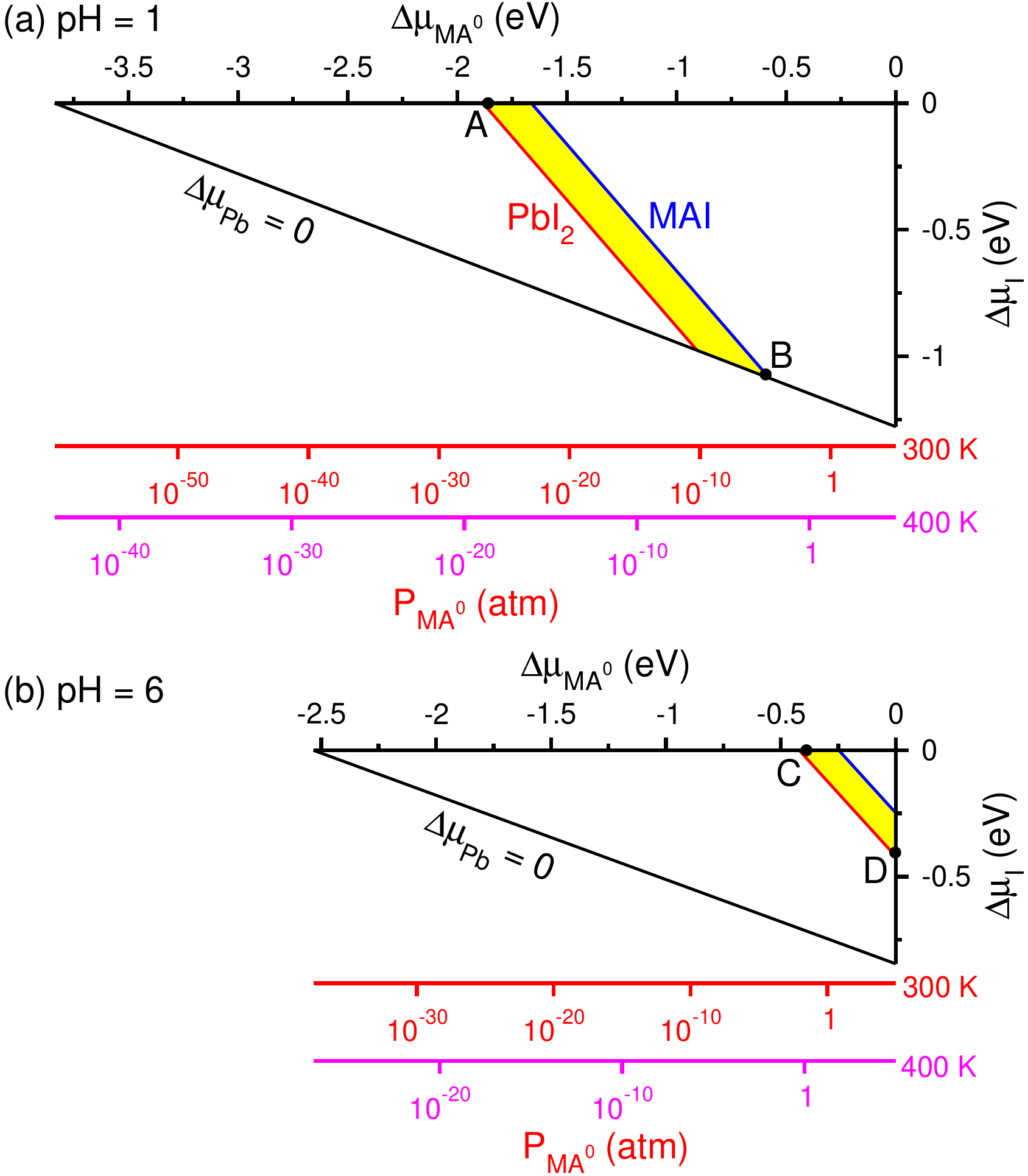}
\caption{Chemical potential triangle formed by \ce{CH3NH2} ($\Delta\mu_{\ce{MA}^0}$: horizontal axis), I ($\Delta\mu_{\ce{I}}$: vertical axis) and Pb ($\Delta\mu_{\ce{Pb}}=0$: oblique line) chemical potentials, with the thermodynamic stable range for equilibrium growth of bulk \ce{MAPbI3} (yellow-colored region) at (a) pH = 1 and (b) pH = 6 conditions.
Red and blue lines indicate the borders for bulk \ce{PbI2} and \ce{MAI} formations, respectively.
The bottom axes present the partial pressure of MA$^0$ gas ($P_{\ce{MA}^0}$) converted from its chemical potential at temperatures of 300 and 400 K.}
\label{fig4}
\end{figure}

\subsection{Surface phase diagram of \ce{MAPbI3} surface}
As the next stage of calculation, we investigated the phase diagram of \ce{MAPbI3} surface under consideration of the different synthesis conditions, which helps understanding of \ce{PbI2} conversion to \ce{MAPbI3}.
To do this, we evaluated the surface formation energies of the low-index \ce{MAPbI3} surfaces with various terminations by using Eq.~\ref{eq4} as varying the chemical potentials of MA$^0$ ($\Delta\mu_{\ce{MA}^0}$) and H ($\Delta\mu_{\ce{H}}$) species.
The range of $\Delta\mu_{\ce{MA}^0}$ was confined by Eq.~\ref{eq6}, giving the relative temperature and pressure conditions by Eq.~\ref{eq10}, while $\Delta\mu_{\ce{H}}$ was controlled to give the pH value between 0 and 14.
For the representative values of pH, we took the two points of pH = 1 and 6, correspondingly $\Delta\mu_{\ce{H}}=-0.26$ and $-1.57$ eV (Table S3), reflecting that the hydroiodic (HI) solution used in the synthesis experiment is strongly acidic.
With these two values of $\Delta\mu_{\ce{H}}$ and the calculated formation enthalpies of the relevant compounds, we determined the lower limits of $\Delta\mu_{\ce{MA}^0}$ as $-3.84$ and $-2.53$ eV at pH = 1 and 6, respectively.
In the same way, the lower limits of iodine chemical potential $\Delta\mu_{\ce{I}}$ were determined to be $-1.28$ and $-0.84$ eV at these pH values.

With the calculated formation enthalpies of \ce{MAPbI3}, \ce{PbI2} and MAI ($-4.10$, $-1.96$, $-1.92$ eV in Table S1), the chemical potentials of MA$^0$ and I for stable \ce{MAPbI3} satisfying Eqs.~\ref{eq7}$-$\ref{eq9} were determined, as shown with the yellow-colored regions in Figure~\ref{fig4}(a) and~\ref{fig4}(b) for pH = 1 and 6, respectively.
The left red- and right blue-colored oblique lines indicate the borders for bulk \ce{PbI2} and \ce{MAI} formations, respectively.
The bottom axes show the partial pressure of MA$^0$ gas converted from its chemical potential at temperatures of 300 and 400 K by using Eqs.~\ref{eq10} and~\ref{eq11} (Table S2).
The narrow shape of the stable region is consistent with the previous calculations regarding the formation of \ce{MAPbI3} from the reaction of \ce{MAI} and \ce{PbI2}~\cite{Yin14apl,Haruyama14jpcl}.
It was found that the chemical potential range for equilibrium growth condition of \ce{MAPbI3} bulk is more confined in the case of pH = 6 than that of pH = 1, indicating that lower pH condition is more favorable for synthesis of \ce{MAPbI3}.
In addition, we found that for the stable region of chemical potentials the partial pressure of MA$^0$ gas ($P_{\ce{MA}^0}$) should be lower than 1 atm at pH = 1 while should be higher at pH = 6.

\begin{figure}[!th]
\centering
\hspace{5pt}
\includegraphics[clip=true,scale=0.15]{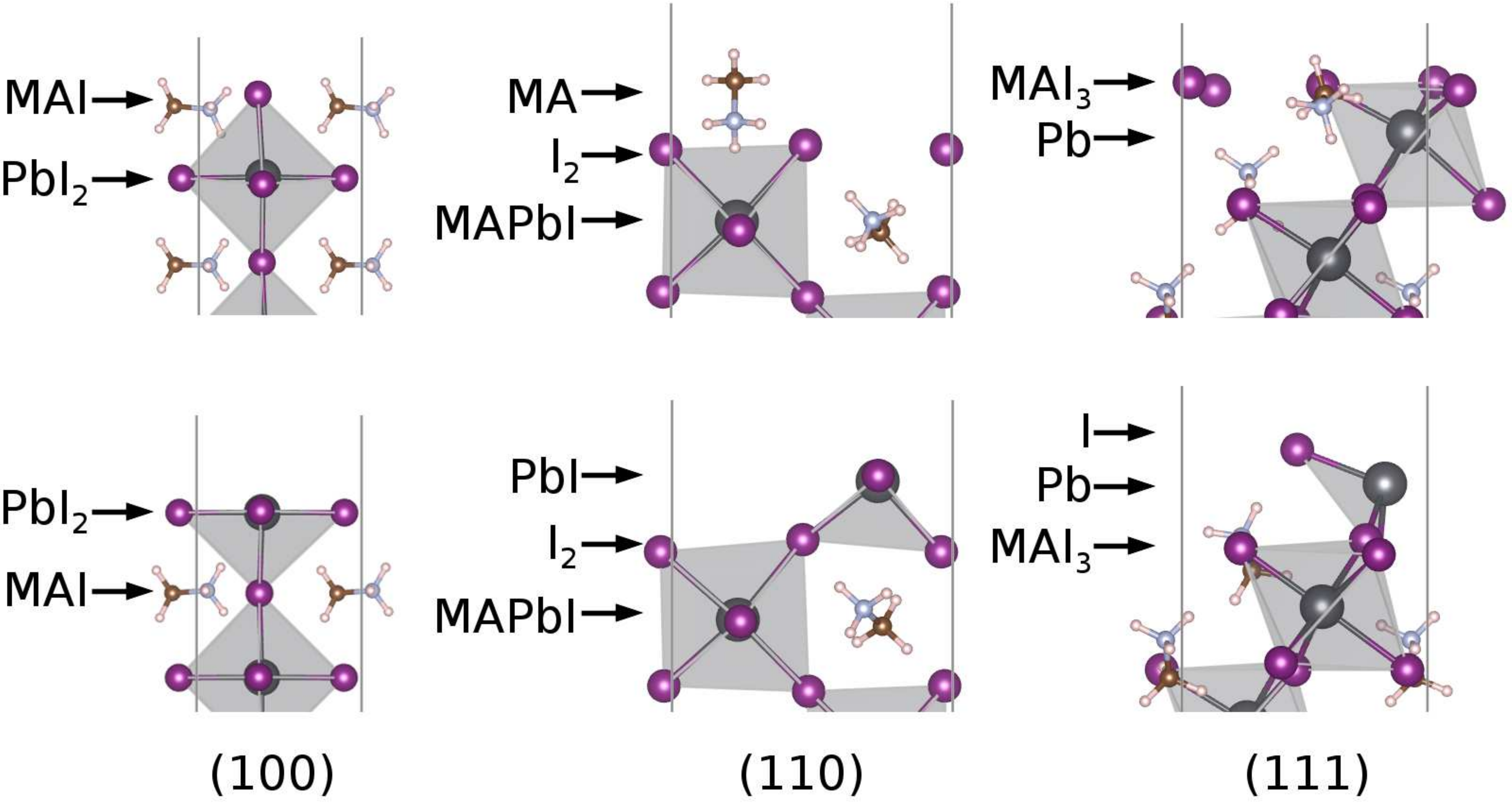}
\caption{Polyhedral view of top surface regions of low-index \ce{MAPbI3} surfaces. For (100) surface (left panel), MAI and \ce{PbI2} terminations with zero net dipole are considered. For (110) surface (middle panel), MAPbI, PbI, \ce{I2} and MA terminations with dipole moment are possible. For (111) surface (right panel), \ce{MAI3}, Pb and I terminations with dipole moment are treated.}
\label{fig5}
\end{figure}
\begin{figure*}[!th]
\centering
\includegraphics[clip=true,scale=0.17]{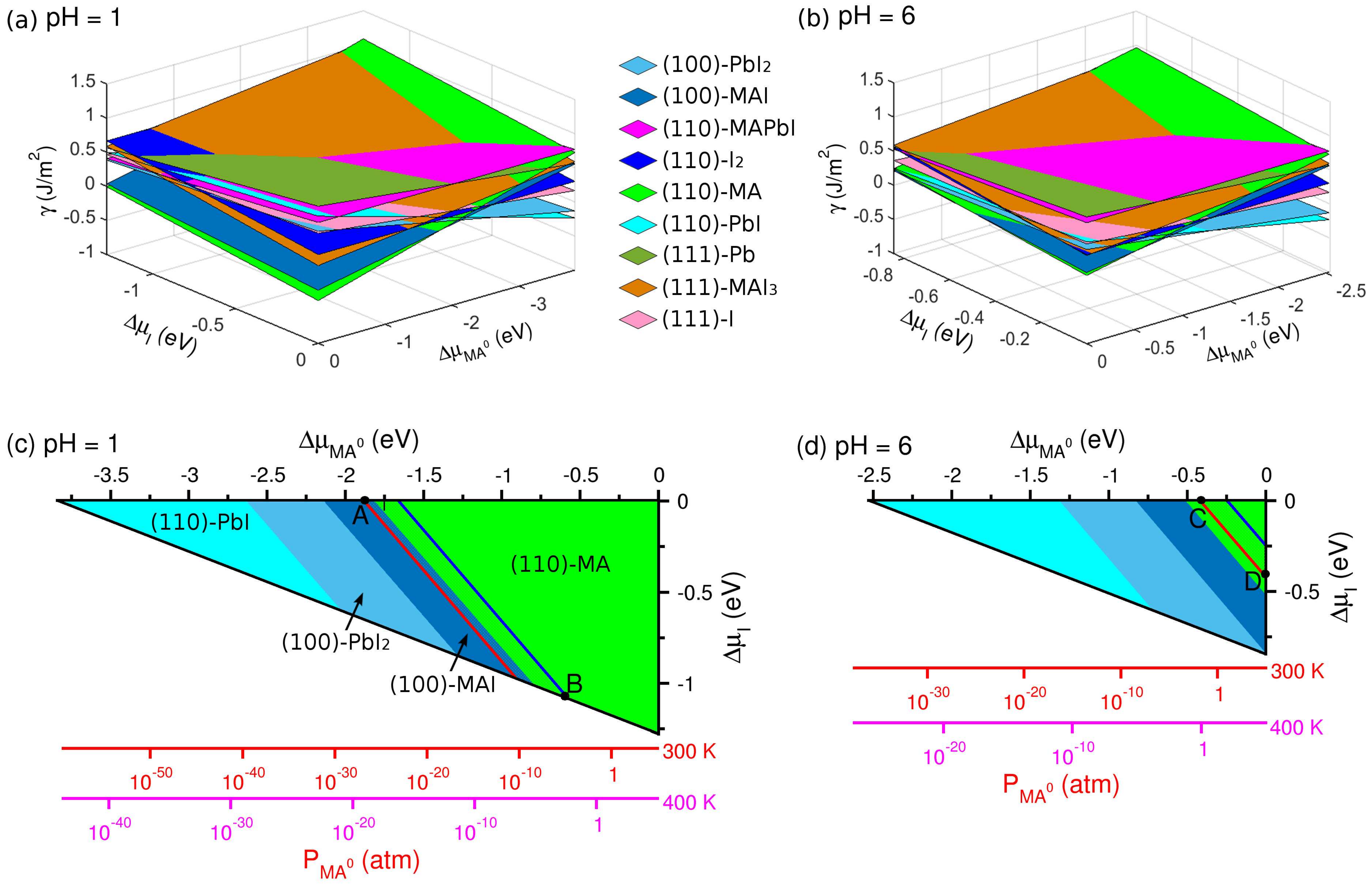}\vspace{0pt}
\caption{Surface formation energies of low index \ce{MAPbI3} surfaces with different terminations at the chemical potential ranges of MA$^0$ and I species under (a) pH = 1 and (b) pH = 6 conditions.
Surface phase diagrams projected onto ($\Delta\mu_{\ce{MA}^0}$, $\Delta\mu_{\ce{I}}$) triangle under (c) pH = 1 and (d) pH = 6 conditions.
The long and narrow regions between red and blue lines indicate the thermodynamically stable regions for equilibrium growth of cubic \ce{MAPbI3} crystal.}
\label{fig6}
\end{figure*}

The morphology of perovskite surface and interface is modified according to the growth condition, which plays a decisive role in photovoltaic performance of PCSs.
In this sense, we drew the surface phase diagram of cubic \ce{MAPbI3} by calculating formation energies of low-index (100), (110) and (111) surfaces with different terminations under aforementioned growth conditions.
We applied two different terminations for the (100) surface, i.e., (100)-\ce{PbI2} and (100)-MAI, both of which are type 1 surfaces with equal anions and cations on each plane according to the Tasker's classification~\cite{Tasker79}.
For the (110) surface, the charged planes stacking alternately produce a dipole moment perpendicular to the surface (type 3), and thus we consider not only two types of typical terminations, (110)-MAPbI and (110)-PbI, but also substantially reconstructed surfaces for stabilization, (110)-\ce{I2} and (110)-MA.
The (111) surface also shows a type 3 stacking sequence, and in the same way, ideal surfaces of (111)-\ce{MAI3} and (111)-Pb along with the modified (111)-I surface were treated in this work.
Figure~\ref{fig5} depicts the different terminations of these low-index surfaces.
To avoid a probable failure in supercell calculations by net dipole moment, we used symmetric slab supercells with the top and bottom surfaces with the same terminations (Figure S7).

Figure~\ref{fig6}(a) and~\ref{fig6}(b) present the three-dimensional graphs showing the surface formation energies of the nine surfaces calculated at the chemical potential ranges of MA$^0$ and I species under pH = 1 and 6 conditions, respectively.
When projecting them onto the ($\Delta\mu_{\ce{MA}^0}$, $\Delta\mu_{\ce{I}}$) plane shown in Figure~\ref{fig4}, two terminated surfaces appeared as stable ones on the plane for each index surface (Figure S8).
In the case of (100) surface, the MAI-terminated surface is stable at the $\ce{MA}^0$-rich condition, while \ce{PbI2}-terminates one is stable at the $\ce{MA}^0$-poor condition.
The thermodynamically stable chemical potential range for the equilibrium growth of \ce{MAPbI3} (the yellow region in Figure~\ref{fig4}) is occupied by the MAI-terminated surface, indicating that this type of surface is a dominant one [Figure S8(a)].
The same result was observed at both pH values, but at pH = 6 condition the \ce{PbI2}-terminated surface appeared further away from the thermodynamically stable region compared with the case of pH = 1.
Among the four terminations of the (110) plane, the MA- and PbI-terminated surfaces are the ground states in the whole chemical potential range.
At both pH values, i.e., for the entire acidic condition range, the MA-terminated surface occupies fully the stable region.
At pH = 1, the (110)-PbI surface is closer to the stable region, indicating that this type of surface might be more probable at low pH conditions than high pH conditions.
Meanwhile, the MAPbI- and \ce{I2}-terminated (110) surfaces have higher formation energies, indicating that these types of surfaces are not likely to appear in synthesized samples, which is consistent with the expectation that these surfaces would be unstable due to being polar.
Similarly, for the (111) surfaces, the polar Pb-terminated surface does not appear on the triangle, but I- and \ce{MAI3}-terminated surfaces appear.
At pH = 1, the reconstructed I-terminated surface is mostly placed on the stable region, whereas at pH = 6 \ce{MAI3}-terminated surface occupies the most part of the stable region.

\begin{table}[!th]
\small
\caption{\label{tab2}The calculated surface formation energies for all types of surfaces considered in this work under MA$^0$-poor/I-rich (point A and C) and MA$^0$-rich/I-poor (point B and D) conditions at pH = 1 (A, B) and 6 (C, D).}
\begin{tabular}{lrrrr}
\hline
& \multicolumn{4}{c}{Surface formation energy $\gamma$ (J/m$^2$)}  \\ \cline{2-5}
Type & Point A & Point B & Point C & Point D \\
\hline
(100)-\ce{PbI2}  & 0.26  & 0.31 & 0.29 & 0.29 \\
(100)-\ce{MAI}  & 0.16  & 0.12 & 0.13 & 0.13 \\
(110)-\ce{MAPbI}  & 0.77  & 0.46 & 0.77 & 0.65 \\
(110)-\ce{I2}  & 0.30  & 0.61 & 0.30 & 0.42 \\
(110)-\ce{MA}  & 0.16  & 0.10 & 0.12 & 0.12 \\
(110)-\ce{PbI}  & 0.33  & 0.39 & 0.37 & 0.37 \\
(111)-\ce{Pb}  & 0.79  & 0.56 & 0.81 & 0.71 \\
(111)-\ce{MAI3}  & 0.36  & 0.59 & 0.34 & 0.44 \\
(111)-\ce{I}  & 0.39  & 0.42 & 0.41 & 0.41  \\
\hline
\end{tabular}	
\end{table}
By combining these, we drew the phase diagram of \ce{MAPbI3} surfaces projected onto ($\Delta\mu_{\ce{MA}^0}$, $\Delta\mu_{\ce{I}}$) triangle, as shown in Figure~\ref{fig6}(c) and ~\ref{fig6}(d) at pH = 1 and 6, respectively.
The chemical potential triangles were found to be occupied by (110)-PbI, (100)-\ce{PbI2}, (100)-MAI, and (110)-MA surfaces in the order of $\Delta\mu_{\ce{MA}^0}$ increasing.
The thermodynamically stable regions for equilibrium growth of cubic \ce{MAPbI3} crystal, with a shape of narrow trapezoid bordered by red and blue oblique lines, were occupied by (100)-MAI and (110)-MA surfaces for pH = 1, while occupied by only (110)-MA surface for pH = 6.
As shown in Figure~\ref{fig6}(c), the (110)-MA and (100)-MAI surfaces coexist on the thermodynamically stable region under pH = 1 condition, indicating the predominance of these surfaces in \ce{MAPbI3} perovskite fabricated from MA$^0$ gas-\ce{PbI2} solid reaction.
On the contrary, at pH = 6, only the (110)-MA surface was found on the whole range of stable region, which is consistent with X-ray diffraction patterns of \ce{MAPbI3} films synthesized from MA$^0$–based reaction route~\cite{Long16natcom}.
It should be noted that the (100)-\ce{PbI2} surface is not so far from the thermodynamically stable range and (110)-PbI surface is stable at only MA$^0$-poor condition being far away from that range.
Table~\ref{tab2} summarizes the calculated surface formation energies of the nine \ce{MAPbI3} surfaces at different points shown in Figure~\ref{fig4}, including A ($\Delta\mu_{\ce{MA}^0}=-1.88$, $\Delta\mu_{\ce{I}}=0$, $\Delta\mu_{\ce{Pb}}=-1.95$ eV), B ($\Delta\mu_{\ce{MA}^0}=-0.58$, $\Delta\mu_{\ce{I}}=-1.09$, $\Delta\mu_{\ce{Pb}}=0$ eV), C ($\Delta\mu_{\ce{MA}^0}=-0.42$, $\Delta\mu_{\ce{I}}=0$, $\Delta\mu_{\ce{Pb}}=-2.11$ eV) and D ($\Delta\mu_{\ce{MA}^0}=0$, $\Delta\mu_{\ce{I}}=-0.42$, $\Delta\mu_{\ce{Pb}}=-1.27$ eV).
In accordance with the phase diagram, the (110)-MA and (100)-MAI surfaces have the lowest formation energies of 0.10-0.16 J/m$^2$ at the four points, then the (100)-\ce{PbI2} surface with 0.26-0.31 J/m$^2$, and the (110)-PbI surface with 0.33-0.39 J/m$^2$.

\begin{figure}[!th]
\centering
\includegraphics[clip=true,scale=0.17]{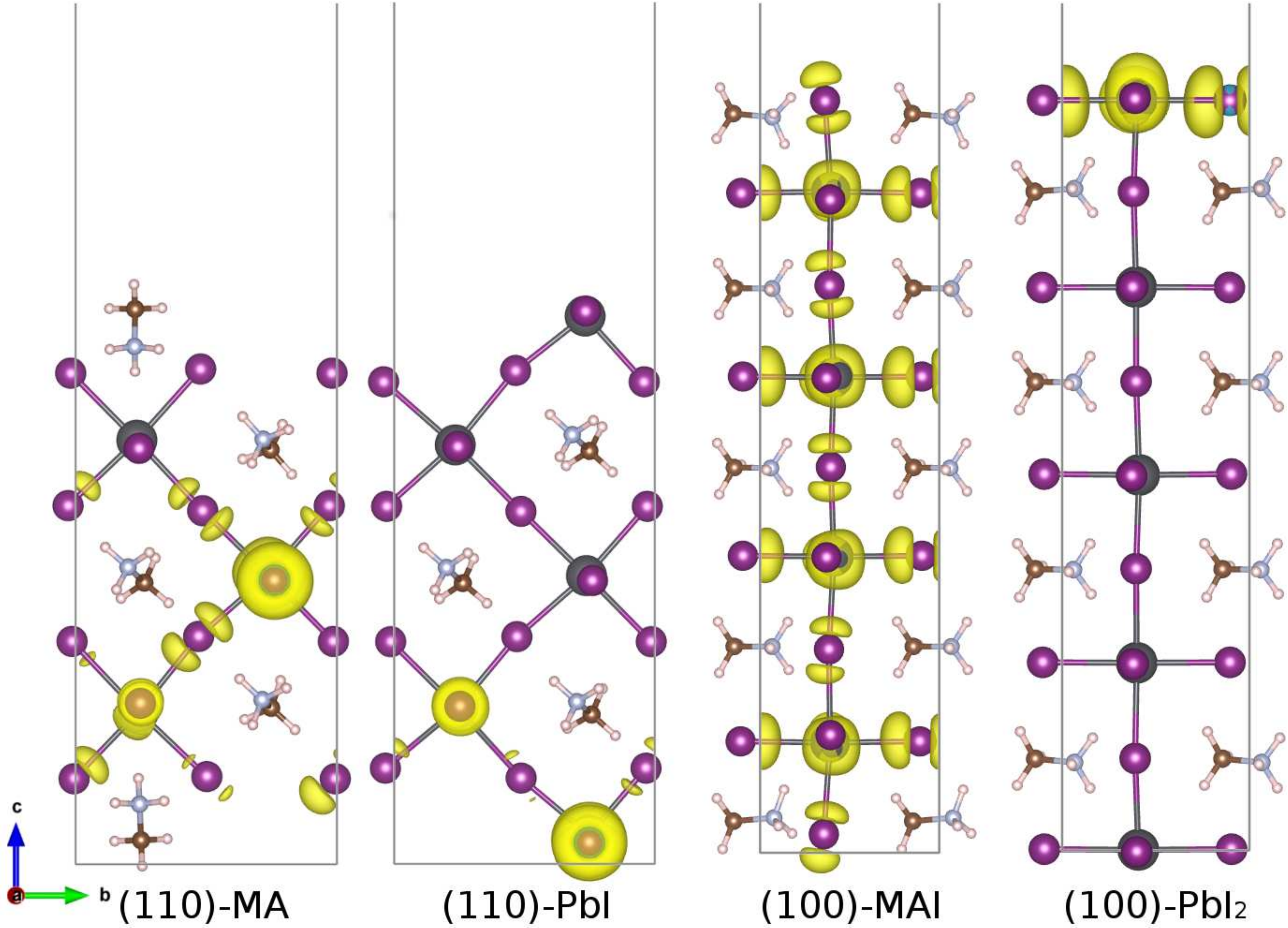}
\caption{Charge distribution of the highest occupied molecular orbitals of (110)-MA, (110)-PbI, (100)-MAI and (100)-\ce{PbI2} surfaces.}
\label{fig7}
\end{figure}
%
We finally analyzed the electronic states of the (110)-MA, (100)-MAI, (100)-\ce{PbI2} and (110)-PbI surfaces.
Figure~\ref{fig7} shows the highest occupied molecular orbitals (HOMOs) of these selected surfaces.
The HOMOs of the (110)-MA surface are found to be distributed around Pb and I atoms inside the slab, while the (100)-MAI termination have the HOMOs spread through the whole slab.
These HOMOs inside the bulk are less effective in transferring the photo-generated holes to the adjoining HTMs, which can affect the incident-photon-to-current efficiency (IPCE). 
In contrast, the HOMOs of the (110)-PbI and (100)-\ce{PbI2} are well localized at the top layer, which are advantageous for enhancing transfer of the holes through those states.

\section{Conclusions}
In this work, we have systematically investigated \ce{PbI2} and cubic \ce{MAPbI3} surfaces by using combined density functional theory and thermodynamic calculations, aiming at a better understanding of precursor effect on perovskite synthesis via solid-gas reaction.
We devised the slab models of \ce{PbI2}(001) and \ce{MAPbI3}(100), (110) and (111) surfaces with perfect and various vacant defect terminations, and calculated their formation energies and adsorption energies of \ce{CH3NH2} molecule on the \ce{PbI2} surfaces under different synthesis conditions of temperature, pressure and pH through chemical potentials of species.
We revealed that the adsorption of \ce{CH3NH2} molecule can be facilitated by including HI or \ce{NH4I} molecule due to lowering adsorption energies and their being negative even at 300 or 400 K through dissociation of this additive and formation of \ce{CH3NH3+} cation or \ce{NH3}-\ce{CH3NH3+} complex, which is advantageous for conversion of \ce{PbI2} to \ce{MAPbI3} via solid-gas reaction.
Moreover, it was found that among nine different types of the \ce{MAPbI3} low-index surfaces, the MA-terminated (110) and MAI-terminated (100) surfaces occupy the thermodynamically stable region for equilibrium synthesis of bulk \ce{MAPbI3} at pH = 1 and 6, being agreed well with the experimental findings.
This work can provide a fundamental understanding of solid-gas reaction for high-crystallinity perovskite synthesis towards high performance perovskite solar cells.

\section*{Acknowledgments}
This work is supported as part of the research project ``Design of New Energy Materials'' (No. 2021-12) by the State Commission of Science and Technology, DPR Korea.
Computations have been performed on the HP Blade System C7000 (HP BL460c) that is owned and managed by Faculty of Materials Science, Kim Il Sung University.

\bibliographystyle{elsarticle-num-names}
\bibliography{Reference}

\end{document}